\documentclass[twocolumn,amsmath,amssymb,prb,showpacs]{revtex4}
\pdfoutput=1
\usepackage{graphicx}
\usepackage{dcolumn}
\usepackage{bm}

\begin{document}

\title{Visualizing Quantum Well State Perturbations of Metallic Thin Films near Stacking Fault Defects}

\author{Alexander A. Khajetoorians}
\altaffiliation[Current Address: ]{Institute of Applied Physics, University Hamburg, Hamburg, Germany}
\author{Gregory A. Fiete}
\author{Chih-Kang Shih}%
 \email{shih@physics.utexas.edu}
\affiliation{%
Department of Physics, The University of Texas at Austin, Austin, Texas, 78712
}%

\begin{abstract}
\noindent We demonstrate that quantum well states (QWS) of thin Pb films are highly perturbed within the proximity of intrinsic film defects.  Scanning Tunneling Spectroscopy (STM/STS) measurements indicate that the energy of these states have a strong distance dependence within 4 nm of the defect with the strongest energetic fluctuations equaling up to 100 meV.  These localized perturbations show large spatially-dependent asymmetries in the LDOS around the defect site for each corresponding quantum well state.  These energetic fluctuations can be described by a simple model which accounts for fluctuations in the confinement potential induced by topographic changes.
\end{abstract}

\pacs{73.20.Hb, 73.61.At, 61.72.Lk, 68.37.Ef}
\maketitle

The curious yet prolific properties of thin Pb quantum films are rooted in its rich thickness-dependent quantum oscillation phenomena resulting from strong quantum size effects (QSE) \cite{Hinch1989,Yeh2000,Zhang1998,Wei2002,Jia2006}.  Robust quantum oscillations of the underlying electronic states can have dramatic consequences related to growth, kinetics, and chemical reactivity \cite{Okamoto2002,Upton2004,Ma2007,Khajetoorians2009}.  Most notably, this system has become a fruitful playground for investigating the interplay between quantum electron confinement and superconductivity.  Initial reports demonstrated that the superconducting transition temperature, $T_{C}$, exhibits a pronounced thickness-dependent oscillation which is well correlated with the beating of the underlying quantum well states (QWS)  \cite{Guo2004,Eom2006,Ozer2006}.

While the initial focus was placed on the correlation between the superconducting gap and QWS, recent studies have raised several questions \cite{Brun2009}.  Surprising observations of a pseudogap have brought into question the role of electron-phonon scattering as it relates to QSE \cite{Wang2009}.  Furthermore, astonishing observations of superconductivity in 2 ML Pb films, which exhibit a strong $T_{C}$ dependence on the underlying interface, independent of the QWS near $E_{F}$, has also raised a question about a simple correlation between QWS and superconductivity \cite{Qin2009}.  In order to understand how the QWS play a role in the above mentioned phenomena as well as other physical phenomena, a deeper and broader understanding of how QSE behave in these films is becoming a quandary that needs to be further addressed \cite{Miller2009}.

\begin{figure}[b]
\centerline{\includegraphics[width= 3.35 in]{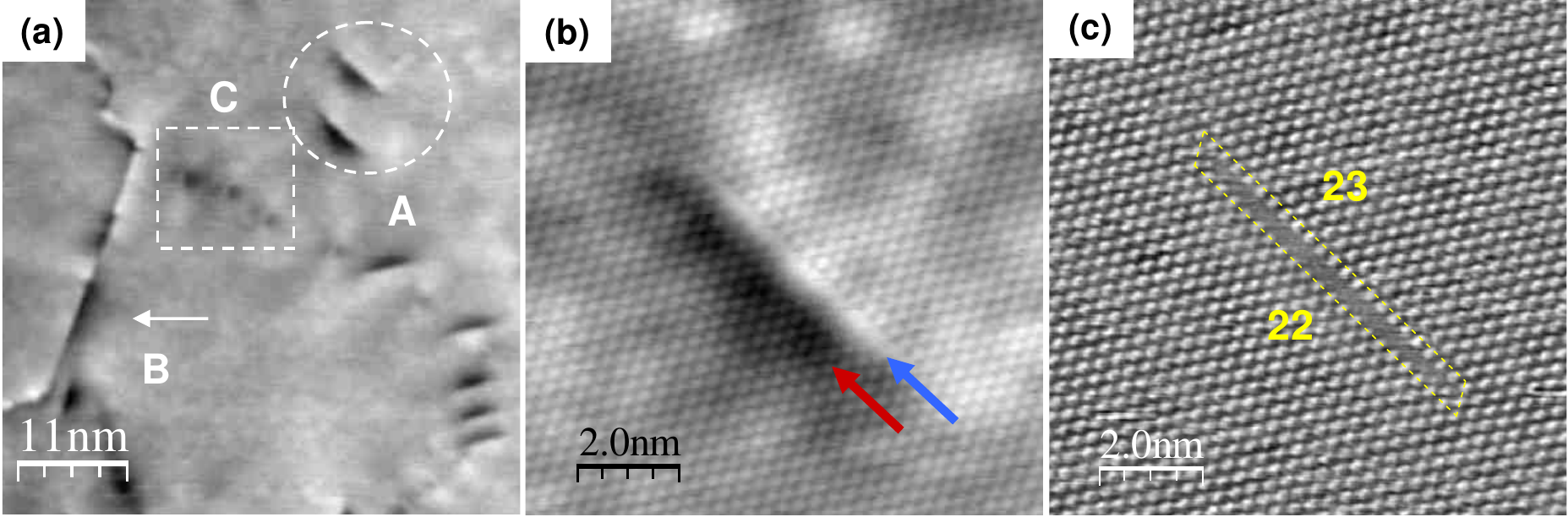}}
\caption{\label{fig:fig1}(color online) (a) STM images of three types of defects (labeled by letter) seen on the surface of globally flat Pb films.  White lines are used to help guide the eye. $V_{sample} = +1.5$ V; $I_{t} = 77$ pA.  (b)-(c) Atomic resolution of a type A defect indicating the ditch region (red arrow) and the spine region (blue arrow), (b) $V_{sample} = -300$ mV; $I_{t} = 25$ pA (c) Laplacian flatting of an atomic image; $V_{sample} = -150$ mV; $I_{t} = 25$ pA; (9 ML $\sqrt{3}$ film).  ).  The number of atoms along the dashed line is indicated.}
\end{figure}

We characterize commonly observed film defects within thin Pb films and demonstrate how the underlying quantum well states (QWS) can dramatically change within the vicinity of these defects. Using scanning tunneling microscopy/spectroscopy (STM/STS), we show that the energetic location of all observed quantum well states can fluctuate as much as 100 meV depending on the lateral location relative to the defect.  Within the vicinity of a defect, these modified states are highly spatially asymmetric resulting from a topographically-induced perturbation to the confinement potential.  We describe these energetic fluctuations within a simple model where the QWS values are interpolated from discrete thickness-dependent values deduced for each film thickness, away from defects, measured by tunneling spectroscopy.

All experiments were carried out using a home built STM operating at $T = 6.5$ K.  PtIr tips and Ir tips were used for all experiments.  STS was taken using a lock-in amplification scheme with a typical modulation frequency of 2 kHz and modulation voltages between $4-10$ mV.  Pb was evaporated from a thermal evaporator and deposited onto either a clean Si$(111)-(7 \times 7)$ surface or $(\sqrt{3} \times \sqrt{3})$-Pb (``$\sqrt{3}$'') interface held near liquid nitrogen temperatures \cite{Ganz1991}.  Globally flat Pb films are formed after annealing cold films near room temperature \cite{Smith1996,Ozer05}.

Pb films exhibit a rather high defect density when utilizing these growth conditions.  Fig \ref{fig:fig1}(a) illustrates an STM image of three different types of intrinsic defects on a Pb film (indicated by letters).  All three defect types appear independent of film thickness (probed up to 25 ML) as well as Si/Pb interface ($7 \times 7$ or $\sqrt{3}$).  Here, we focus only on the properties of type A defects.

Type A defects are oriented along one of three $\langle 0 \overline{1} 1 \rangle$ directions.  Each type A defect has a characteristic semicircular region of lower contrast, referred to as the ``ditch'', which is preferentially located along one side of the defect for a given $\langle 0 \overline{1} 1 \rangle$ orientation.  These ditches have a typical corrugation of $0.5 - 0.9$ \AA\ and radius of $2.5-4$ nm.  For a given $\langle 0 \overline{1} 1 \rangle$ defect orientation, there is an abrupt interface between the ditch and a localized region of higher contrast, which we refer to as the ``spine," that extends along the dislocation line.  Both spine and ditch regions maintain similar contrast at all observed imaging conditions regardless of which part of the unit cell is imaged.  Therefore, the resultant contrast is topographic in origin.

\begin{figure}[t]
\centerline{\includegraphics[width= 3.35 in]{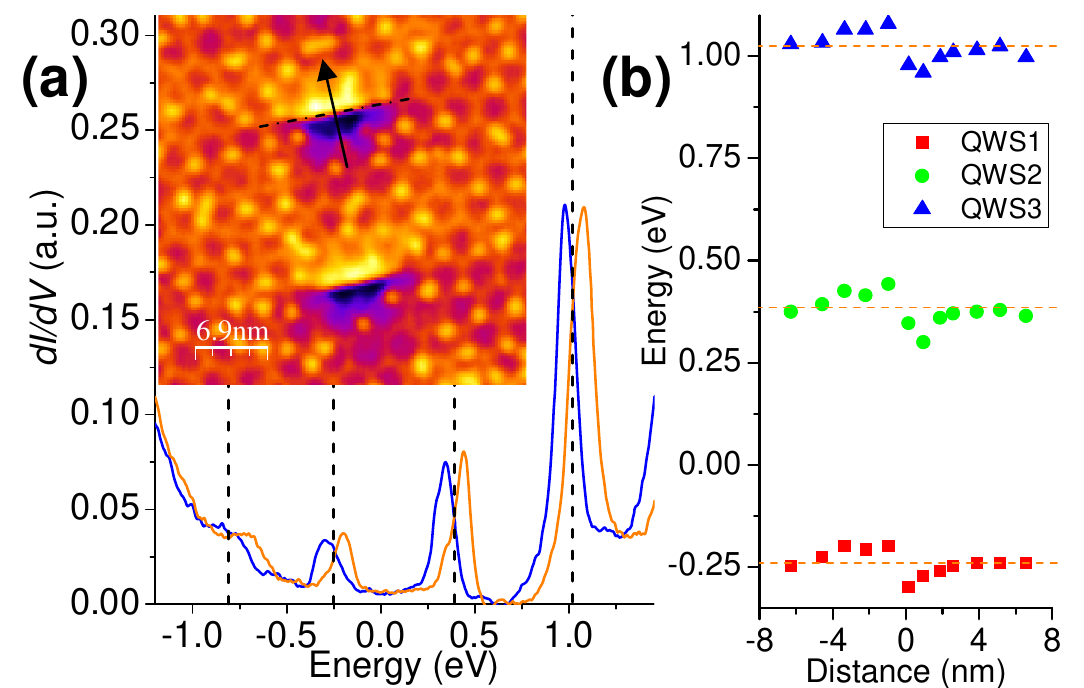}}
\caption{\label{fig:fig2}(color online) (inset) Topography image of type A defects referencing spectroscopy taken in (a)-(b) $V_{sample} = +321$ mV; $I_{t} = 100$ pA.  (a) STS taken in a ditch region (blue) and in a spine region (orange) of an 18 ML Pb film (b) QWS energies derived from spatially resolved STS taken along a perpendicularly bisection line across the interfacial region between ditch and spine region of the defect indicated in (inset) by the arrow.}
\end{figure}

Fig \ref{fig:fig1}(b) illustrates atomically resolved images of a type A defect.  A clear $1 \times 1$ pattern corresponding to an fcc close-packed surface with a periodicity of 0.35 nm can be seen.  As shown in Fig \ref{fig:fig1}(c), a Burger's circuit taken around the defect exhibits one extra atom along the spine side of the defect corresponding to a Burgers vector $\mathbf{b} = \frac{a}{2}[01\overline{1}]$.  While this is equivalent to a single perfect dislocation, a dislocation of this type cannot account for experimental observations.  According to Frank's Energy Criterion, a perfect dislocation with $\mathbf{b}_{1}$ will dissociate into partial dislocations $\mathbf{b}_{2}$ and $\mathbf{b}_{3}$ if $(b_{1}^{2} > b_{2}^{2} + b_{3}^{2})$ \cite{Hirthbook}.  For a perfect dislocation of the given form, this results in two Shockley partial dislocations of the form $\mathbf{b}_{S} = \frac{a}{6} \langle 1 \overline{1}  \overline{2} \rangle$.  This dissociation produces a stacking fault on the surface plane, but will not produce a spine or a ditch.  A possibility which will produce a ditch-like feature is a stacking fault which is situated along a close-packed plane off-axis (e.g. $(11\overline{1})$) from the surface normal (e.g. $70.5^{\circ}$) which extends into the surface.  The displaced atoms beneath the surface can produce an overall contrast change which would result in the appearance of a ditch but maintain atomic order as viewed from the surface.  We believe this is responsible for the ditch feature.  However, favorable dissociations of a perfect dislocation along an off-axis close packed plane or a Frank partial ($\mathbf{b}_{F} = \frac{a}{3}[11\overline{1}]$) would alone not conserve the measured Burgers vector.

Combinations of dislocations can dissociate and interact thereby producing features similar to both ditch and spine \cite{Schmid92,Christiansen2002}.  For the PtNi(111) surface, similar ditches have been attributed to dissociations of a subsurface dislocation network which produces an off-axis stacking fault along a close-packed plane which is bounded by other partial dislocations.  STM images of these types of ditches are very similar to what is observed for a type A defect.  Furthermore, it was demonstrated that the depth of the dislocation network is related to the overall depth of the ditch and shape of the ditch as seen with STM.  This can explain the variation in the corrugation of our ditch.  However, the exact structure which produces these PtNi ditches are not accompanied by a vacancy row (spine) and the ditches are not hemispherically shaped.  Nevertheless, this again suggests that multiple dislocations, with subsurface components, are involved in this structure.

Other possible dislocations where two perfect dislocations along different close-packed planes dissociate and interact to form a sessile dislocation (stair-rod dislocation/Lomer-Cottrell lock) are intriguing options \cite{Christiansen2002}.  These structures have subtle differences with what is observed in our case.  Without further experimental data or simulations which can produce information about participating subsurface dislocations, it is not directly evident what exact dislocation(s)/dissociations beneath the surface produce this defect.
\begin{figure}[t]
\centerline{\includegraphics[width= 3.35 in]{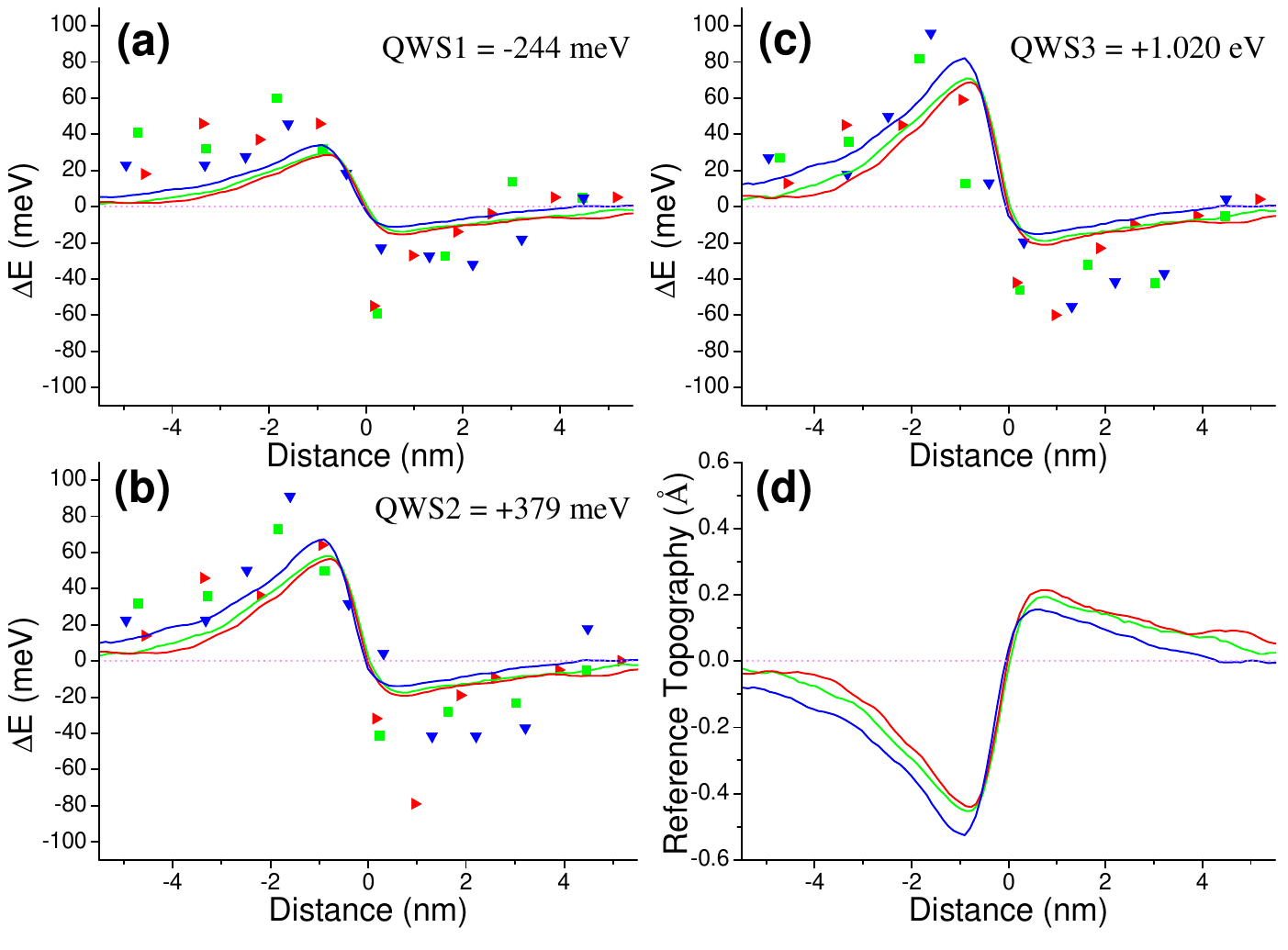}}
\caption{\label{fig:fig3}(color online) (a) - (c) Relative shifts in QWS energies derived from STS for 3 separate defects (symbols). $\Delta E$ refers to the energy location of a QWS subtracted from the state energy away from the defect.  Each color refers to a specific defect.  Solid lines refer to the calculated energy shifts from the interpolation model.  (d) Reference line profiles for the three defects used to calculate the energy shifts in (a)-(c) (reference indicated in Fig 2(a)).}
\end{figure}

\begin{figure*}[t]
\centerline{\includegraphics[width= 5.8 in]{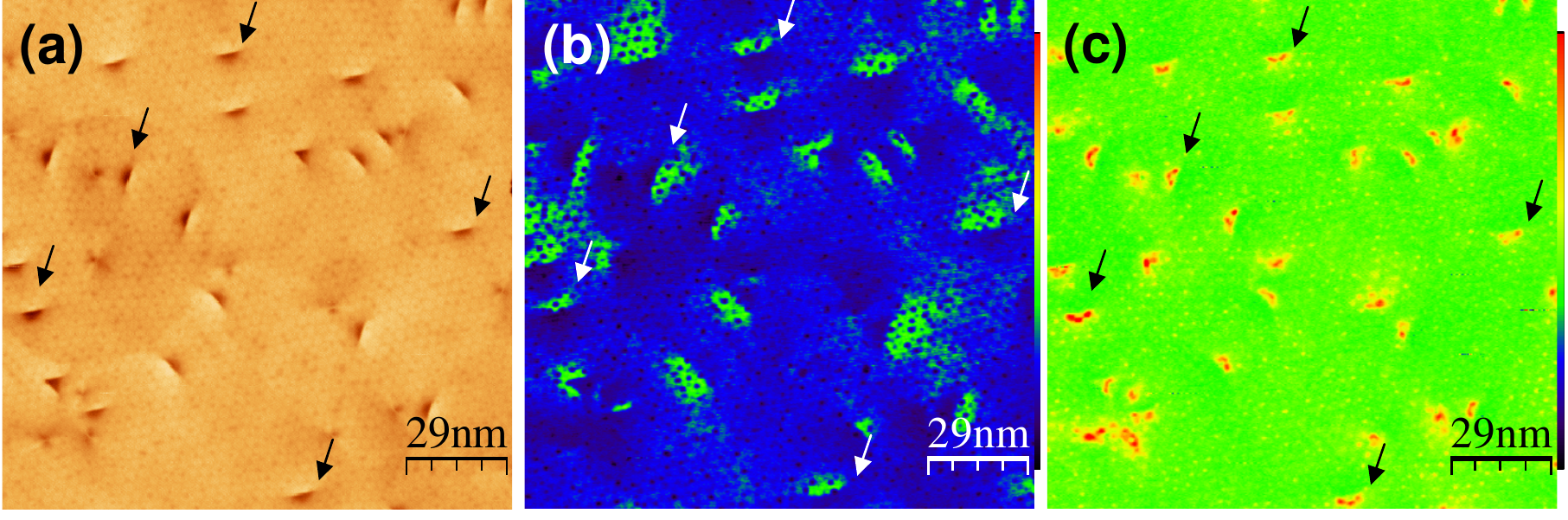}}
\caption{\label{fig:fig4}(color online) Topography and constant-current dI/dV images ($V_{mod} = 4$ mV).  (a) Topography $V_{sample} = +470$ mV; $I_{t} = 95$ pA. (b) Constant-current $dI/dV$ image $V_{sample} = +321$ mV; $I_{t} = 100$ pA, (scale: $0.8-2.2$ V) (c) Constant-current $dI/dV$ image $V_{sample} = +470$ mV; $I_{t} = 95$ pA, (scale: $0-1.4$ V).  Arrows indicate specific defects which show different characteristic spatial distributions.  All Pb films were 18 ML thick grown on a $\sqrt{3}$ Pb-Si interface.}
\end{figure*}

Scanning tunneling spectroscopy (STS) within the vicinity of these defects reveals well-defined states resulting from quantum confinement \cite{Eom2006,Khajetoorians2009}.  However, there is a vast variation in QWS energies within the vicinity of a defect depending on the probed location (Fig \ref{fig:fig2}(a)).  QWS exhibit lower overall energies when probed in the spine (orange) and higher overall energies when probed in the ditch (blue).  There is a clear spatially-dependent trend in the energetic shifts of the QWS (Fig \ref{fig:fig2}(b)) which shows the strongest modulations near the interfacial region.  Each color/symbol corresponds to a particular state for the given film thickness.  Less than 4 nm from the interface, specifically within the ditch region of the defect, there is a monotonic increase in all state energies as the orthogonal distance decreases between the probed region and the interface.  At the interface, there is an abrupt decrease in all the state energies which is far below the average energy value for each state for a given thickness away from the defect.  These values monotonically increase as the orthogonal distance from the spine region increases until all state energies are restored to their typical value away from the defect.  All energetic shifts are localized within a radius of $\approx 4$ nm from the center of the interfacial region.

Fig \ref{fig:fig3}(a)-(c) illustrate the measured energy fluctuations across three different type A defects on a 18 ML film.  Here $\Delta E$ represents the state energy at the given location subtracted from the average state value of the film away from the defect.  Negative distance values reference the ditch region (Fig \ref{fig:fig3}(d)).  Energetic shifts within the ditch region are persistently larger in magnitude, for each state, as compared to the spine region.  Furthermore, the shifting behavior shows a weak state dependence, especially for the ditch region, where the overall shifting at a given point increases with absolute energy.  All type A defects show the same energetic shifting behavior independent of the defect orientation and underlying interface.  All spectroscopic points for a given defect were taken using similar stabilization conditions in order to avoid Stark shift effects created by the tip \cite{Limot2003}.

The overall change in state energy can be understood within the context of a simple perturbation to the confinement potential.  By interpolating the $l$-th QWS $E_{l,n}$ energy value for a given discrete thickness $n$, $dE_{l,n}/dz$ can be extracted for non-integer thickness values where $z$ represents any real-valued thickness \cite{Khajetoorians2009}.  The simulated changes in state energies can be calculated by multiplying the deviations in topography near the defect, $\Delta z$, by $dE_{l,n}/dz$ \cite{note3}.  The solid lines in Fig \ref{fig:fig3}(a)-(c) represent the calculated changes in energy for each defect assuming the absolute height away from the defect is the zero reference (Fig \ref{fig:fig3}(d)).  This simple model reproduces the observed experimental trends such as the state dependence of the energetic shifting as well as the overall order of magnitude of the changes in state energy in the defect regions.

The overall enhancement of the LDOS near a defect site shows a highly non-homogenous spatial distribution at particular energies.  Fig \ref{fig:fig4} illustrates constant-current $dI/dV$ images and the correlated topography image taken below ($E = +321$ meV; $\Delta E \approx -60 $ meV) and above ($E = +470$ meV; $\Delta E \approx +90$ meV) (Fig a given QWS for a typical 18 ML $\sqrt{3}$ film.  Strong enhancement of the LDOS can be seen near type A defects both below and above the given state energy ($E_{QWS} = +379$ meV).  For downward energy shifts, the enhanced LDOS in the spine region shows a more isotropic and localized distribution. The enhancement spans the length of the interfacial region and is localized $\approx 2$ nm within the spine region.  For the ditch region, the LDOS exhibits similar enhancements.  However, unlike the spine region, there are pronounced geometrical features with strong intensities in the ditch region of each defect (indicated by arrows).  The number of lobes and the overall shape of these features varies depending on the diameter and apparent depth of the ditch.  However, the geometrical shape for a given defect is nearly identical \emph{above all probed states}.  This is unlike the case where localized lobe features, attributed to quantum confinement, change as a function of energy \cite{Maltezopoulos2003,Folsch2004}.  These lobes have a characteristic diameter between $1.5 - 4$ nm.  There is some indication that these lobes may be quantized in multiples of $\lambda_{F}/2$.  We believe these spatial distributions may be understood within the context of electron scattering within proximity of the dislocations of the defect.  However, this requires a more detailed model than the one presented.

A description of the resultant dipole moment emanating from a defect site was first predicted by Smoluchowski \cite{Smoluchowski1941}.  Within the context of the jellium model, a dipole moment can result from an abrupt change in morphology which leads to a smearing in the charge density on the order of $\lambda_{F}/2$.  However, a more complete picture of how this resultant polarization is produced requires a detailed account of the modified bonding of both atoms and vacancies within the vicinity of the defect \cite{Muller2006}.  Few measurements have been able to detect this polarization \cite{Jia1998,Park2005}.  For these cases, the measured fields can be described by a step-edge model.  Descriptions of more complex defects have eluded theoretical modeling because of the intensive resources required.  In context of this work, an overall modulation of the QWS naturally results in a modulation of the work function which should produce a local polarization.  The field should be localized on the order of $\lambda_{F}/2$ within the defect along the off-axis close packed plane of the defect where $\lambda_{F}/2 \approx 0.5$ nm \cite{Zhang2005}.  This local field may further perturb the QWS within the vicinity of the defect.  This may explain why the downward shifts in the spine region, within 1 nm of the interfacial region, are much larger than predicted from our model. Nevertheless, from spectroscopy alone, we cannot distinguish between QWS fluctuations resulting from topographical perturbations and those produced by dipolar contributions.

In conclusion, we demonstrate that within the vicinity of intrinsic defects there exists large modulations in the QWS ($\Delta E \approx 100$ meV).  These fluctuations can be described by a topographically-induced perturbation to the local confinement potential.  LDOS maps reveal that at the corresponding local QWS energy near the defect sites that the LDOS has a rather complex distribution. These large fluctuations in the overall LDOS should be considered as these Pb systems become increasingly utilized for a variety of experiments involving kinetics, surface chemistry, and superconductivity.

The authors would like to thank Wenguang Zhu and Paulo Ferreira for fruitful discussions.  The authors would like to acknowledge funding from ARO: W911NF-09-1-0527, NSF-DMR-0906025, NSF-DGE-0549417, and Welch Foundation: F-1672.

\bibliography{Citations-updated}

\begin{thebibliography}{10}
\bibitem{note3}%
  \BibitemOpen
  \bibinfo {note} {We linearly inteprolate the energy values between each
  discrete thicknesses therefore making the $dE/dz$ constant}%
  \bibAnnoteFile{NoStop}{note3}%
\end{thebibliography}

\end{document}